\documentclass[letterpaper, 10 pt, conference]{ieeeconf}  

\IEEEoverridecommandlockouts                              

\usepackage[english]{babel}
\usepackage{cite}
\usepackage{amsmath,amssymb,amsfonts}
\usepackage{algorithmic}
\usepackage{graphicx}
\usepackage{textcomp}
\usepackage{amsmath}
\usepackage{xcolor}
\usepackage{units}
\usepackage{booktabs}
\usepackage{comment}
\usepackage{lettrine}
\newcommand{\flexbrac}[1]{\if\relax\detokenize{#1}\relax \else (#1) \fi}
\newcommand{\flexcomma}[1]{\if\relax\detokenize{#1}\relax \else ,#1 \fi}

\newcommand{\abs}[1]{\lvert #1 \rvert}

\newcommand{\cA}{\mathcal{A}}

\newcommand{\cC}{\mathcal{C}}

\newcommand{\cG}{\mathcal{G}}

\newcommand{\cI}{\mathcal{I}}

\newcommand{\cK}{\mathcal{K}}

\newcommand{\cR}{\mathcal{R}}

\newcommand{\cT}{\mathcal{T}}

\newcommand{\cV}{\mathcal{V}}

\usepackage{amsthm}

\newtheorem{theorem}{Theorem}[section]

\newcounter{thm}
\newtheorem{prob}[thm]{Problem}
\def\BibTeX{{\rm B\kern-.05em{\sc i\kern-.025em b}\kern-.08em
		T\kern-.1667em\lower.7ex\hbox{E}\kern-.125emX}}
	
\newif\ifmargincomments 
\margincommentstrue
	
\ifmargincomments

\newcommand{\msmargin}[2]{{\color{red}#1}\marginpar{\color{red}\raggedright\footnotesize [MaS]:\\ #2}}
\else

\newcommand{\msmargin}[2]{#1}
	\fi

\title{\LARGE \bf
Cost-optimal Fleet Management Strategies for\\ Solar-electric Autonomous Mobility-on-Demand Systems
}

\author{Fabio Paparella$^1$ , Theo Hofman$^1$ and Mauro Salazar$^1$
\thanks{$^{1}$The authors are with the Control Systems Technology section, Department of Mechanical Engineering, Eindhoven University of Technology, 5600 MB Eindhoven, The Netherlands
        {\tt\small \{f.paparella,t.hofman,m.r.u.salazar\}@tue.nl}}%
}

\begin{document}

\maketitle

\begin{abstract}
This paper studies mobility systems that incorporate a substantial solar energy component, generated not only on the ground, but also through solar roofs installed on vehicles, directly covering a portion of their energy consumption.
In particular, we focus on Solar-electric Autonomous Mobility-on-Demand systems, whereby solar-electric autonomous vehicles provide on-demand mobility, and optimize their operation in terms of serving passenger requests, charging and vehicle-to-grid (V2G) operations.
We model this fleet management problem via directed acyclic graphs and parse it as a mixed-integer linear program that can be solved using off-the-shelf solvers.
We showcase our framework in a case study of Gold Coast, Australia, analyzing the fleet's optimal operation while accounting for electricity price fluctuations resulting from a significant integration of solar power in the total energy mix.
We demonstrate that using a solar-electric fleet can reduce the total cost of operation of the fleet by 10-15\% compared to an electric-only counterpart.
Finally, we show that for V2G operations using vehicles with a larger battery size can significantly lower the operational costs of the fleet, overcompensating its higher energy consumption by trading larger volumes of energy and even accruing profits.
\end{abstract}





\section{Introduction}
\lettrine{A}{s the rise} of Mobility-as-a-Service (MaaS) paradigms has been revolutionizing transportation, this sector is one of the few where emissions are still increasing~\cite{EPA2018}.
In this context, recent advances in autonomous driving, and electric powertrain and vehicle design could provide pathways to a more sustainable car-based mobility:
Combining centrally controlled autonomous vehicles with powertrain electrification could enable the deployment of Electric Autonomous Mobility-on-Demand (E-AMoD) systems, whereby a fleet of electric vehicles provides on-demand mobility.
The possibility of centrally controlling the fleet provides opportunities not only in optimizing the vehicles' routes and schedules, but also in jointly optimizing their charging and vehicle-to-grid (V2G) activities, profitably trading energy in markets where electricity prices are significantly varied, for instance as a countermeasure to the so-called \textit{duck curve}~\cite{Denholm2015}, typical of energy mixes with a large share of solar energy.
Moreover, the deployment of such fleets allows and calls for vehicles that are tailored to the specific application.
In this particular paper, we consider highly-efficient electric vehicles that can be equipped with a solar rooftop directly covering a portion of their energy demand, and study their deployment within Solar-electric Autonomous Mobility-on-Demand (S-AMoD) systems.
Against this background, the paper studies fleet operational strategies that are optimized to satisfy user demands, whilst profitably performing price-driven charging and V2G activities.

\textit{Related Literature:}
This paper pertains to the research streams of mobility-on-demand operation and solar-electric vehicles, which we review in the following.
Multiple approaches to model and control AMoD systems are available: Two examples are the vehicle routing problem (VRP) ~\cite{YaoChenEtAl2021b,PavoneFrazzoli2010,PavoneBisnikEtAl2007,TothVigo2014,Laporte1992,PsaraftisWenEtAl2016,PillacGendreauEtAl2013} and multi-commodity network flow models~\cite{SpieserTreleavenEtAl2014,IglesiasRossiEtAl2018,PaparellaPedrosoEtAl2023}.
Both methods are flexible and allow for the implementation of a wide range of constraints and objectives. Some examples are congestion-aware routing~\cite{RossiZhangEtAl2017, Wollenstein-BetechHoushmandEtAl2020}, and  VRP with time windows and heterogeneous fleet composition~\cite{HoffAnderssonEtAl2010b,BaltussenGouthamEtAl2023}. 
When considering E-AMoD systems, vehicle coordination and charging algorithms have been studied in~\cite{BoewingSchifferEtAl2020}, whilst also accounting for infrastructure placement in~\cite{LukeSalazarEtAl2021,PaparellaChauhanEtAl2023}.
In this context, the interaction between large-fleet AMoD systems and power distribution networks has been extensively studied in~\cite{RossiIglesiasEtAl2018b,EstandiaSchifferEtAl2021}, whereas in our recent work~\cite{PaparellaHofmanEtAl2022} we have optimized the operations of an E-AMoD system jointly with its fleet composition in terms of number of vehicles and battery size of the individual vehicles. Nevertheless, none of these papers focuses on solar-electric systems, and their design and operations within price-varying energy markets. 

Focusing on solar-electric vehicles, some studies have investigated them from a technological standpoint~\cite{Connors2007,WamborikarSinha2010,WaseemSherwaniEtAl2019}, but mostly focusing on vehicular aspects, and without exploring deployment and operational questions.

In conclusion, there are no available studies that investigate S-AMoD systems in terms of deployment and operations, optimizing their routing, charging and V2G activities.

\textit{Statement of Contributions:}
This paper presents a modeling framework to optimize the operations of an S-AMoD fleet, and carry out design studies by devising a directed acyclic graph (DAG) model describing the fleet routing, charging and V2G activities.
Next, we leverage such a model to optimize the fleet management strategies via mixed-integer linear programming, with the objective to maximize the profit of the fleet operator. 
Finally, we showcase our framework on a case study of Gold Coast, Australia, where we assess the benefits of S-AMoD in terms of operational costs, and compare different vehicular designs.

\textit{Organization:} The remainder of this paper is structured as follows: Section \ref{sec:Form} introduces the S-AMoD model optimization framework, whilst Section~\ref{sec:Res} details our real-world case study of the city of Gold Coast, Australia, and the results obtained.
Finally, Section~\ref{Sec:Concl} draws the conclusions and provides an outlook on future research.

\section{Problem Formulation}\label{sec:Form}
In this section, we model the vehicle routing, charging and V2G operations problem via DAGs, extending our preliminary model~\cite{PaparellaHofmanEtAl2022} to fully capture V2G activities, solar-electric vehicles and time-varying electricity prices.

\subsection{Directed Acyclic Graph Model}
We define a transportation network as $\mathcal{G'} = (\mathcal{V'},\mathcal{A'})$, where the set of arcs $\cA'$ are roads and the set of nodes $\cV'$ are intersections between roads. We denote a set of charging stations $\cC \subset \cV'$ in the transportation network, with station $c\in \cC$. We also define travel request $i \in \mathcal{I}:=\{1,2...,I\}$ as $r_i = (o_i,d_i,t_i) \in \cV'\times \cV' \times \cT$ being a travel request on network $\cG'$ from origin $o_i$ to destination $d_i$ at time $t_i\in \cT$. 
First, we expand the set of travel requests to $\mathcal{I}^+:=\{0,1,2...,I,I+1\}$ where the first and last requests represent a fixed location (deposit, parking spot), so that vehicles start and conclude their tasks at a pre-defined point.
Then, we construct a DAG, $\mathcal{G}=(\mathcal{V},\mathcal{A})$, containing deposits and transportation requests.
Each arc $(i,j) \in \mathcal{A}$ represents the fastest path from the destination of $r_i$, $d_i$, to the origin of $r_j$, $o_j$ and it is characterized by travel time  $t^\mathrm{fp}_{ij}$ and distance $d^\mathrm{fp}_{ij}$, respectively. If $i=j$, we denote with $t^\mathrm{fp}_{ii}$ the time of the fastest path to serve request $i$.
Last, we define a set of vehicles $\cK :=\{1,2...,K\}$ with vehicle $k\in\cK$.

\subsection{Operational Constraints}
We define a transition matrix $X \in \{0,1\}^{(I+1) \times (I+1) \times K}$, where $X_{ij}^k=1$ if vehicle $k$ serves demand $i$ and then demand $j$, and zero otherwise. We introduce the tensor $S \in \{0,1\}^{(I+1) \times (I+1) \times K \times C}$ to account for charging and V2G activities. If vehicle $k$ in-between $d_i$ and $o_j$ goes to charging station $c$, $S_{ijc}^k$ is set to one, otherwise it is set to zero.
In the same way, we quantify the amount of energy dis(charged) with element $C_{ijc}^k$ of the corresponding charging tensor $C \in \cR^{(I+1) \times (I+1) \times K \times C}$. If the element is positive, energy is withdrawn from the grid; if negative, energy is injected into the grid, meaning that vehicle $k$ performs V2G.   

Travel, deviation and charging times can be pre-computed via standard shortest path algorithms, thus we can directly eliminate all the unfeasible transitions from a time perspective. In particular, $t^\mathrm{ava}_{ij} := t_j-t_i + t_{ii}^\mathrm{fp}$ is the available time between the destination of $i$ and the origin of request $j$. Thus, we can pre-compute the upper bounds of the tensors as
\begin{equation}\label{eq:boundX}
	X_{ij}^k\leq 
	\begin{cases}
		1 & \text{if} \; t_{ij}^{\mathrm{fp}} \leq t^\mathrm{ava}_{ij}\\
		0 & \text{otherwise},
	\end{cases}  
\end{equation}
\begin{equation}
	{S}_{ijc}^k \leq 
	\begin{cases}
		1 & \text{if} \; t_{ij}^\mathrm{fp} +  \Delta T_{ijc}^\mathrm{go2S} \leq t^\mathrm{ava}_{ij}\\
		0 & \text{otherwise},
	\end{cases}   
\end{equation}
\begin{equation}\label{eq:Cmax}
	\abs {{C}_{ijc}^k } \leq
	\begin{cases}
		\hat{C}_{ijc}^k & \text{if} \; t_{ij}^\mathrm{fp} +  \Delta T_{ijc}^\mathrm{go2S} \leq t^\mathrm{ava}_{ij}\\
		0 & \text{otherwise},
	\end{cases}   
\end{equation}
with $i,j\in\cI^+,\;c \in \cC,\;k\in\cK$. The extra time needed to go to charging station $c$ during transition $ij$ is $ \Delta T_{ijc}^\mathrm{go2S}$, while the upper bound  \mbox{$\hat{C}_{ijc}^k=( t^\mathrm{ava}_{ij} -t_{ij}^\mathrm{fp} -  \Delta T_{ijc}^\mathrm{go2S})P_\mathrm{ch}$} is the energy (dis)charged if all the available time were  used to (dis)charge at the given charging power $P_\mathrm{ch}$.
To enforce that each request can be served at most once, we define the transitions constraints
\begin{equation}\label{eq:maxonce1}
	\sum_{i\in \cI^+,k\in \cK}  X_{ij}^k+ \sum_{k\in \cK} f_{j}^k \leq 1\;\;\;\;\forall j\in \cI^+,
\end{equation} 
\begin{equation}\label{eq:maxonce2}
	\sum_{j\in \cI^+,k\in \cK}  X_{ij}^k+ \sum_{k\in \cK} l_{i}^k \leq 1 \;\;\;\;\forall i\in \cI^+.
\end{equation} 
To guarantee continuity of the schedules---meaning that if a vehicle performs transition $ij$, its next one will start from $j$---we impose
\begin{equation}\label{eq:conti}
	\sum_{i\in \cI^+} X_{ij}^k - \sum_{l\in \cI^+} X_{jl}^k = f_j^k+l_j^k\;\;\;\;\forall j\in \cI^+,\;\forall k\in \cK.
\end{equation}
We enforce that vehicle $k$ can charge between demand $i$ and $j$ at station $c$, only if it serves both requests,
\begin{equation}\label{eq:SOnlyIfX}
	\sum_{c\in \cC}S_{ijc}^k \leq  X_{ij}^k \;\;\;\;\forall i,j\in \cI^+,\;\forall k\in \cK,
\end{equation}
and goes to station $c$ in-between,
\begin{equation}\label{COnlyIfS}
	\abs{C^k_{ijc}} \leq \hat{C}_{ijc}^k  \cdot S_{ijc}^k\;\;\;\;\forall i,j\in \cI^+,\;\forall c\in \cC,\;\forall k\in \cK.
\end{equation}
Last, $f$ and $l$ are two parameters to set the initial and final position of the vehicles in the deposit, so that \mbox{$f_j^k,l_j^k=0 \;\forall i,j \in \cI,\forall k\in\cK$} and $f_0^k=l_{I+1}^k=1 \; \forall k\in\cK$.
\subsection{Energy Constraints}
In this section we introduce the energy constraints for the vehicles. Defining the energy level of vehicle $k$ after serving demand $j$ as $e^k_{j}$, we formulate an energy balance at each node of the DAG as
\begin{align}
	e^k_{j} =  e^k_{i} - E_{ij}^k &+ \sum_{c\in \cC} C_{ijc}^k+ C^{\mathrm{sol}}_{ij} \nonumber \\ 
	&\forall i,j \in \cI^+,\forall k \in \cK |X_{ij}^k=1,\label{eq:EbalJ}
\end{align}
where $e^k_{j}$ is the summation of $e_i^k$, the energy of vehicle $k$ at $d_i$, the energy required to transition from $i$ to $j$ and serve $j$ $E_{ij}^k$, the energy charged at any charging station, and the solar energy charged from the roof during transition $ij$ $C^\mathrm{sol}_{ij}$. The transition energy is defined as
\begin{equation}\label{eq:EbalTrans}
	E^k_{ij} = 
	\begin{cases}
		d^\mathrm{fp}_{ij}\cdot \bar{E}_\mathrm{con} \quad \forall i,j \in \cI^+,\forall k \in\cK |\sum_{c\in \cC} S_{ijc}^k=0
	
		\\
		\begin{split}
			(d^\mathrm{fp}_{ij} + \Delta& d_{ijc}^\mathrm{go2S} ) \cdot \bar{E}_\mathrm{con} \\
			&\forall i,j \in \cI^+,\forall c \in\cC,\forall k \in\cK |S_{ijc}^k=1,
		\end{split}
	\end{cases}
\end{equation}
where $\bar{E}_\mathrm{con}$ is the consumption per unit distance of the vehicles, $d^\mathrm{fp}_{ij}$ is the distance of the fastest path from $d_i$ to $o_j$, and $\Delta d_{ijc}^\mathrm{go2S}$ is the additional distance driven to pass by station~$c$. Finally, we bound the state of energy of each vehicle to be within its battery capacity as
\begin{equation}
	0 \leq e^k_{j} \leq E_ \mathrm{b}^\mathrm{max} \; \;\forall j \in \cI^+,\;\forall k\in\cK,
\end{equation}
and set the initial and final battery level to a pre-defined value $E_\mathrm{b}^0$ as
\begin{equation}\label{eq:boundTime}
	e^k_{0}= e^k_{I+1} = E_\mathrm{b}^0\;\;\forall k\in\cK.
\end{equation}

\subsection{Objective}
In this paper, we set the optimization objective as the maximization of the profit accrued by the fleet.
The two terms that influence it are the cost of operation and the revenues generated by serving requests.
Formulated as a cost-minimization function, the objective is then
\begin{equation}\label{eq:obj}
	J=  \sum_{i,j\in \cI} p_{ij}^\mathrm{el} \cdot \sum_{k \in \cK} \sum_{c \in \cC} C^k_{ijc}  - \sum_{i \in \cI^+} b_\mathrm{r}^i \cdot p_i,
\end{equation}
where $p_{ij}^\mathrm{el}$ is the price of electricity during transition $ij$, approximated by its average between $t_i+ t^{\mathrm{fp}}_{ii}$ and $t_j$, the binary variable $b_\mathrm{r}^i$ indicates whether request $i$ is being served, and $p_i$ is the revenue generated from it.

Then the maximum-profit operation problem for an S-AMoD fleet is defined as follows:

\begin{prob}[Optimal S-AMoD Fleet Management]\label{prob:main}
Given a set of transportation requests $\cI$, the operations maximizing the total profit of the S-AMoD system result from
	\begin{equation*}
		\begin{aligned}
			\min\; &J\\
			\mathrm{s.t. }\; &\eqref{eq:boundX}-\eqref{eq:obj}. 
		\end{aligned}
	\end{equation*}
\end{prob}
Problem~\ref{prob:main} is a mixed integer linear program that can be solved with global optimality guarantees by off-the-shelf optimization algorithms.

\subsection{Discussion}
A few comments are in order. First,  we consider travel times to be known. This assumption is in order if the routing of the fleet does not affect travel time, i.e., traffic congestion effects are exogenous.
Second, we assume the electricity prices to be known in advance during the entire day, given the predictability of the macro-trend of the price fluctuations.
Third, given the offline analysis and design purpose of the work, we assume the travel requests to be known in advance. For real-time operational purposes, it would be possible to implement an online version of this framework in a receding horizon fashion.
Fourth, Problem~\ref{prob:main} is NP-hard. Hence it is not possible to solve it for large instances. However, by solving stochastically sampled scenarios, we can draw a sub-optimal, but more conservative solution of the original problem~\cite{PaparellaHofmanEtAl2022}. 
Fifth, we assume the amount of energy traded by the S-AMoD fleet not to influence the energy prices, as is the case for medium to large fleets: For instance, for the case study presented in Section~\ref{sec:Res} below, the considered fleet of 800 vehicles trades less than 0.2\% of the overall electric energy trading volume of the region.
We leave the study of the coupling between the two systems with significantly higher S-AMoD penetration rates to future research.
Sixth, we assume spatially-homogeneous weather conditions, so that the solar energy harvested by each vehicle $C^\mathrm{sol}_{ij}$ depends on the time of day only.
Finally, we aim at maximizing the profit generated by the operator, yet our framework can readily accommodate different cost models, which we leave to future research endeavors.

\section{Results}\label{sec:Res}
In this section, we compare an S-AMoD and E-AMoD system in terms of operational strategies and costs, and how these are influenced by the electricity price fluctuations caused by a high proportion of solar power in the energy mix, and by different vehicle design choices.
\vspace{-0.1cm}
\subsection{Case Study of Gold Coast, Australia}
We present a case study conducted in the city of Gold Coast, Queensland, Australia, using data on the road network and travel requests obtained from Transportation Networks for Research~\cite{ResearchCoreTeam}. 
Our analysis focuses on over 30,000 travel requests served by a fleet of 800 vehicles, comparing the performance of both S-AMoD and E-AMoD systems. The simulation is conducted for both winter and summer, as these periods exhibit distinct energy price and solar radiation profiles throughout the day, allowing us to observe variations in fleet operation. As a base vehicle for the S-AMoD system, we consider the highly-efficient solar-electric Lightyear~0~\cite{Lightyear}. To ensure a fair comparison, for the E-AMoD fleet we employ the same base vehicle with disabled solar roofs. We gather data on battery size $E_\mathrm{b}^\mathrm{max} = \unit[60]{kWh}$ and energy consumption $\bar{E}_\mathrm{con} = \unit[0.12]{kWh/km}$ from the website of the Dutch Car Company Lightyear~\cite{Lightyear}, indicating that, over a sunny day, its rooftop produces approximately \unit[6]{kWh} during summer, and \unit[5]{kWh} during winter. We set the charging power to $P_\mathrm{ch}=\unit[22]{kW}$. 
Finally, we take the cost-models representing the revenues generated when serving the requests from~\cite{Uber:2018}.
Considering its combinatorial complexity, to solve Problem~\ref{prob:main}, we solve smaller randomly sampled scenarios multiple times, similar to~\cite{PaparellaHofmanEtAl2022}, whereby in each sample 200 daily requests are served by 5 vehicles.
We parse the problems with Yalmip~\cite{Loefberg2004} and solve them with Gurobi~10.1~\cite{GurobiOptimization2021}. 
\begin{figure}[ht!]
	\centering
	\includegraphics[width=0.97\columnwidth]{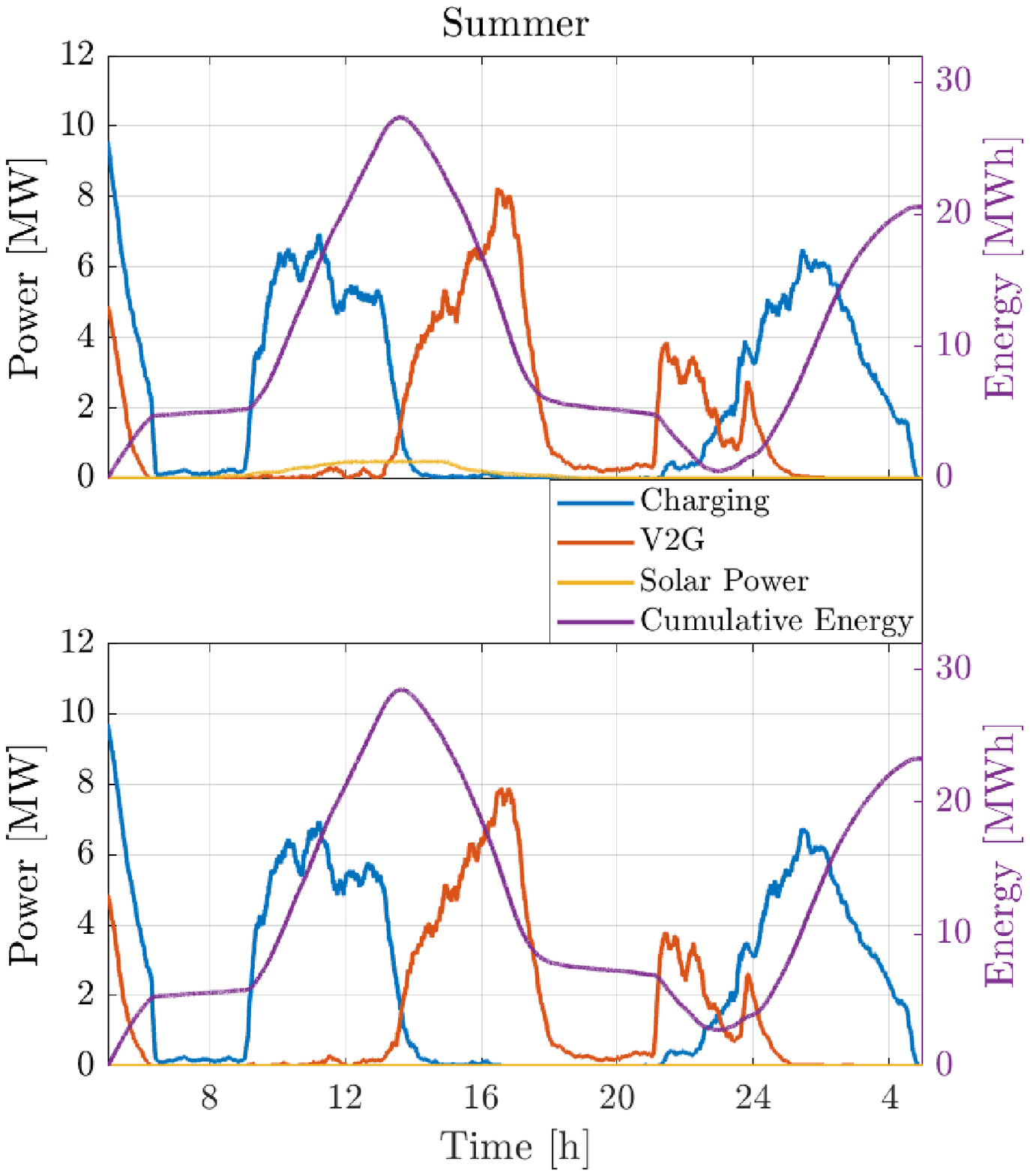}
	\includegraphics[width=1\columnwidth]{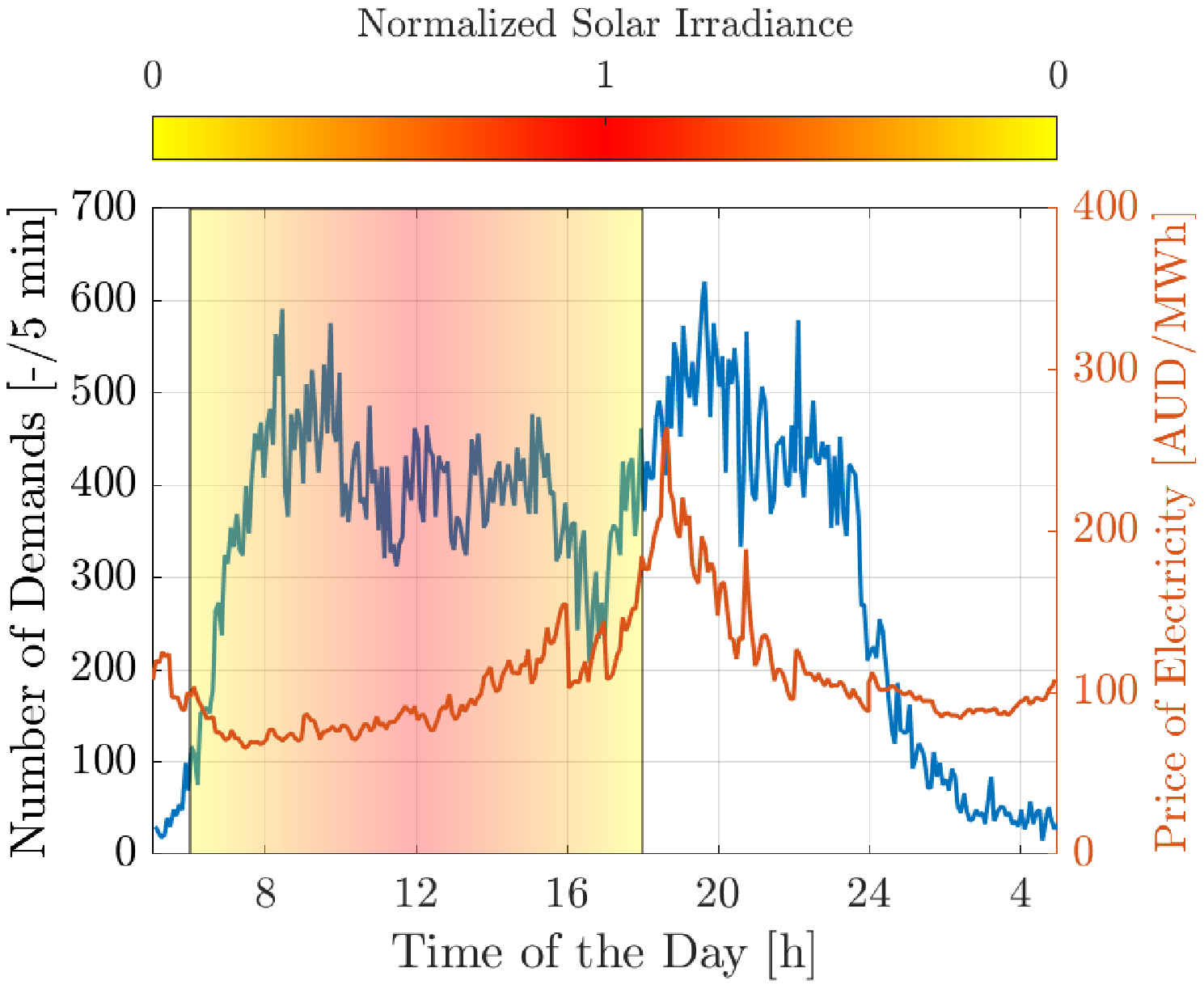}
	
	\caption{The top two figures show the results during summer for an S-AMoD and E-AMoD fleet, respectively, in terms of charging power (blue), V2G power (red), solar power generated by the roof (yellow), and overall cumulative energy consumption (purple). Interestingly, the bottom figure shows the average electricity price in summer 2022 in Queensland (Courtesy of Australian Energy Market Operator) and the number of demands every 5 minutes. The colored area denotes the solar irradiance.} 
	\label{fig:PowerS}
	\vspace{-0.4cm}
\end{figure}
\begin{figure}[ht!]
	\centering
	\includegraphics[width=0.99\columnwidth]{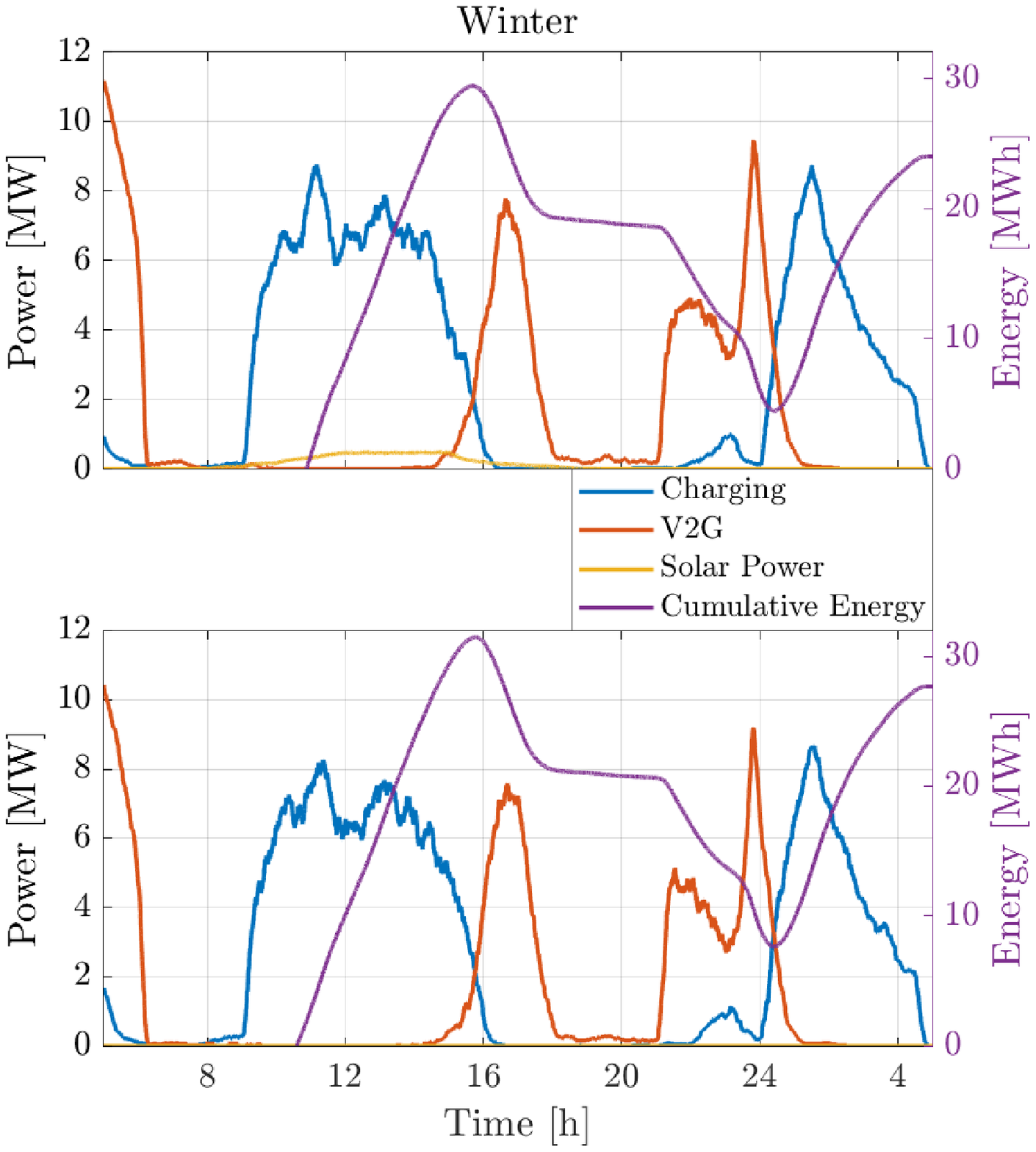}
	\includegraphics[width=1\columnwidth]{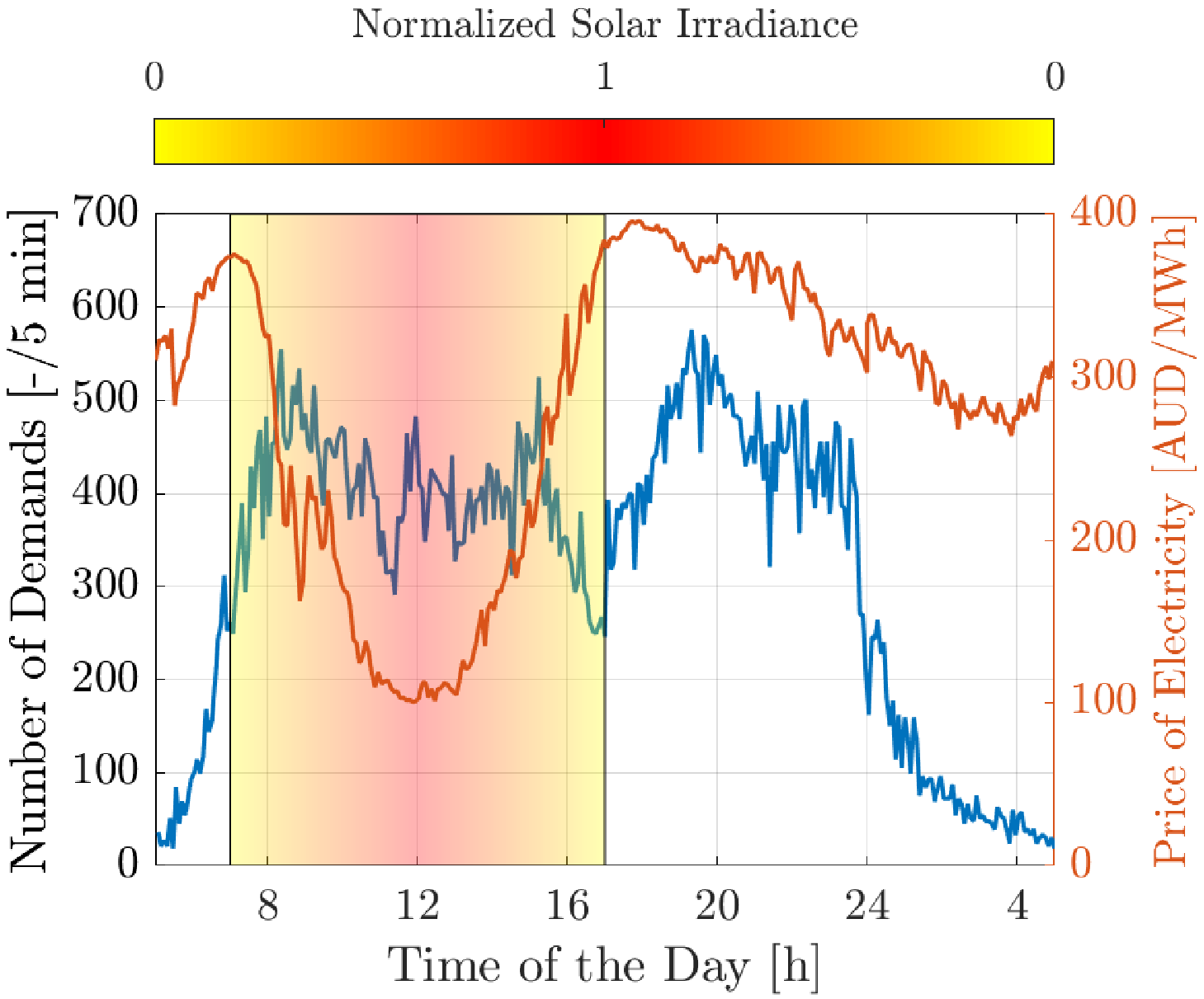}
	\caption{The figures show a simulation during winter for an S-AMoD and E-AMoD systems. During winter, the operator fully recharges the fleet during midday and during night, and conveniently discharges during the two peaks at 7:00 and 18:00.    } 
	\label{fig:PowerW}
		\vspace{-0.4cm}
\end{figure}

Figs.~\ref{fig:PowerS} and \ref{fig:PowerW} show the power withdrawn from and injected to the grid, as well as the overall energy consumed by the S-AMoD and E-AMoD fleet during one day in summer and winter, respectively.
It is noteworthy that the winter and summer case studies exhibit striking similarities: Despite significant price variations, the operator can employ structurally similar strategies: Prioritize charging during midday and nighttime when electricity prices and the number of travel requests are low, and make use of V2G when electricity prices are high and the number of travel requests is low. This approach allows for the assignment of more vehicles in the fleet with discharging or charging tasks during periods of low demand, thereby increasing the operator's overall profitability. In all the four cases, the revenues generated by serving all the travel requests were the same, approximately \unit[265]{kAUD}. This result highlights that, independently of every other factors, the task with the highest priority of the fleet is to serve requests, whilst charging and V2G operations are of secondary importance, given that they account for less than 10\% of the objective function considered.

\begin{figure}[t]
	\centering
	\includegraphics[width=1\columnwidth,trim={0 0 30 5},clip]{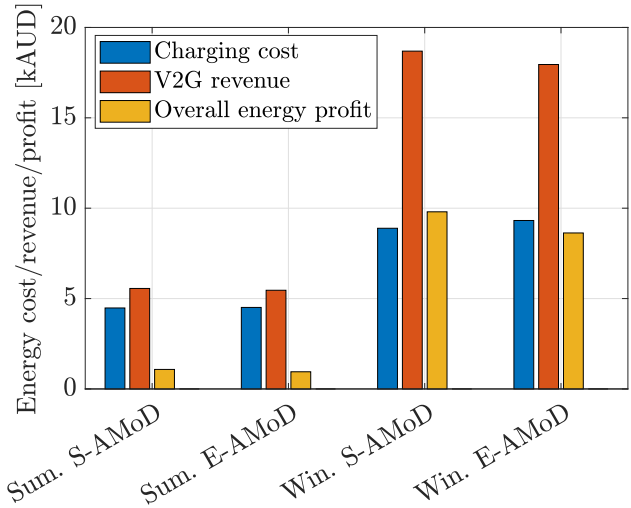}	
	\caption{Charging cost, V2G revenue, and overall profit generated by trading energy during summer and winter for S-AMoD and E-AMoD systems. The revenue generated by serving travel requests is \unit[265]{kAUD} in all four cases.} 
	\label{fig:bar1}
	\vspace{-0.3cm}
\end{figure}

Considering the dis(charging) aspect, Fig.~\ref{fig:bar1} reveals that trading energy is more profitable in winter compared to summer, thereby incentivizing the fleet to undertake more trips for charging and discharging purposes, which, in turn, results in a higher energy consumption.
Then, comparing S-AMoD with E-AMoD, we note that, on the one hand, the operational strategies are very much the same: Indeed, with an average daily solar panel production of \unit[5-6]{kWh}, each vehicle saves approximately \unit[10-15]{min} of charging time. However, this time-saving is not significant since the fleet has ample daily time-windows available for charging. Furthermore, it is crucial to observe that the surplus energy from the solar roofs coincides with periods of abundant solar production in the energy mix, resulting in lower electricity prices.
Consequently, the fleet operator does not benefit significantly from the additional driving range provided by the solar roofs, as it aligns with the fleet's natural recharging periods.
On the other hand, the energy profitability, excluding the part related to the travel requests that coincides in all scenarios, is approximately 10\% higher due to the energy generated by the solar roof.
\vspace{-0.1cm}
\subsection{Comparing Fleets with Different Battery Size}
In this section, we examine the impact of the solar roof for fleets with smaller battery sizes, where the surplus energy can potentially have a more pronounced effect. We conduct a similar case study as the previous section, focusing on summer, and compare S-AMoD and E-AMoD fleets with battery sizes of \unit[20]{kWh} and \unit[40]{kWh}, respectively. We capture the lower energy consumption per unit distance caused by the smaller battery sizes, by accounting for the lower weight of the vehicles in their energy consumption models as shown in~\cite{PaparellaHofmanEtAl2022} and~\cite[Ch.~2]{GuzzellaSciarretta2007}. 
Furthermore, considering charging-levels standards, we set the charging powers to $P_\mathrm{ch} = \unit[8]{kW}$ and \unit[12]{kW}, respectively, ensuring a similar C-rate between the different types of vehicles~\cite{GuzzellaSciarretta2007}.

\begin{figure}[t]
	\centering
	\includegraphics[width=1\columnwidth,trim={20 8 12 12},clip]{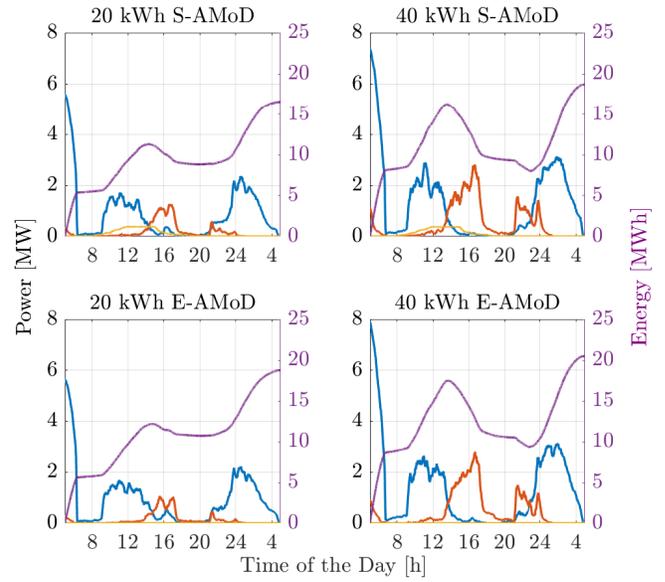}	
	\caption{Charging and discharging operations during summer. The four cases are: S-AMoD and E-AMoD, both  with 20 and 40 kWh battery pack. The meaning of the lines-color is the same as in Figs.~\ref{fig:PowerS} and \ref{fig:PowerW}. } 
	\label{fig:comp}
\end{figure}

\begin{figure}[t]
	\centering
	\includegraphics[width=1\columnwidth,trim={0 0 30 5},clip]{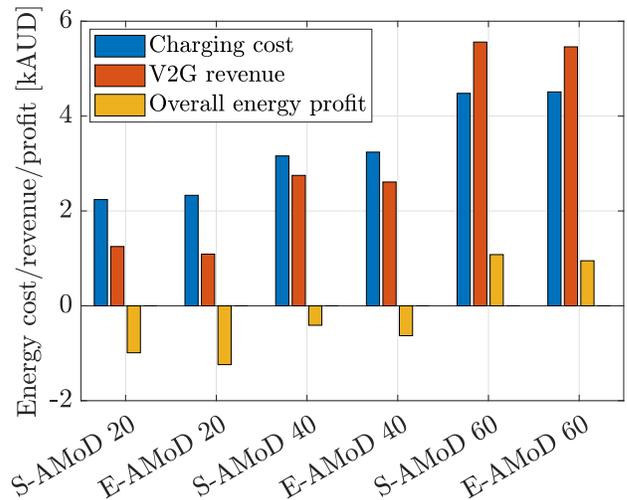}	
	\caption{Charging cost, V2G revenue, and overall profit generated by trading energy during summer for S-AMoD and E-AMoD systems with vehicles equipped with a 20, 40 and \unit[60]{kWh} battery. Note that the last two groups of bars correspond to the first two of Fig.~\ref{fig:bar1}.} 
	\label{fig:bar2}
	\vspace{-0.3cm}
\end{figure}

Fig.~\ref{fig:comp} illustrates the discharging and charging operations in the four cases from which we learn that the lower the battery capacity, the lower the overall amount of energy exchanged.
The E-AMoD fleets equipped with \unit[20 and 40]{kWh} batteries achieve revenues from serving travel requests equal to \unit[240]{kAUD}, 9\%  lower compared to the \unit[60]{kWh} case study.
This is due to the longer time spent charging due to the lower charging power.
Notably, for solar-electric fleets, this gap goes down to 6\% (\unit[245]{kAUD}), showing that, for smaller vehicles, the solar-energy surplus not only positively benefit the energy costs shown in Fig.~\ref{fig:bar2}, but also the overall number of requests that can be served.

Overall, in terms of battery size, this case study shows that when V2G is possible, it is financially more advantageous to employ a fleet with a larger battery size, albeit its higher energy consumption, due to higher trading volumes and the ability to use a stronger charging power.
Moreover, the convenience of having a solar rooftop increases for smaller battery sizes, whereby it can increasingly counter-balance the lower power of the charging stations, and allow a higher number of requests to be served.
Finally, we recall that it would be also important to consider the amortized fixed costs to purchase the fleet, which may be higher for larger batteries, and battery lifetime.
Whilst the present paper is focused on operational costs only, we leave such a comprehensive investigation of the fleet's total cost of ownership to future research.


\section{Conclusions}\label{Sec:Concl}
In this paper, we introduced a framework for optimizing the operation of a Solar-electric  Autonomous Mobility-on-Demand (S-AMoD) fleet with vehicle-to-grid capabilities.
To this end, we leveraged directed acyclic graphs and mixed-integer linear programming to solve the maximum-profit fleet management problem w.r.t.\ the fleet operator point of view.
We showcased our framework through real-world case studies conducted in Gold Coast, Australia, during summer and winter, showing that actively trading energy could lower operational costs for the fleet.
Furthermore, we found that whilst there were no significant operational differences in terms of strategy between a standard electric fleet and a solar-electric fleet, the S-AMoD system exhibited a 10-15\% improvement in operational costs due to a reduction in energy consumption, resulting in additional reductions in $\mathrm{CO}_2$ emissions and overall environmental impact.
Finally, we demonstrated that deploying a fleet with a larger battery size is more profitable when V2G operations are considered, even when accounting for the resulting higher energy consumption of the vehicles.

Moving forward, several extensions to this work are worth exploring.
First, incorporating ride-sharing into the framework would be a natural progression~\cite{PaparellaPedrosoEtAl2023}, as well as studying intermodal settings where transportation requests are served jointly with public transit and active modes~\cite{Wollenstein-BetechSalazarEtAl2021}.
Second, we would like to study the solutions stemming from different cost-functions, such as environmental impact and battery-lifetime-aware total cost of ownership.
Finally, it would be worthwhile developing tailored solution algorithms to solve the optimization problems presented, as well as deriving implementable online control schemes.
\section{Acknowledgments}
We thank Dr. I. New, L. Pedroso and J.P. Bertucci for proofreading this paper.
This publication is part of the project NEON with project number 17628 of the research program Crossover which is (partly) financed by the Dutch Research Council (NWO).
\vspace{-0.1cm}


\bibliographystyle{IEEEtran}
\bibliography{Bibliography/main.bib,SML_papers.bib}

\end{document}